\documentclass[
twocolumn,
english,
aps,
pra,
longbibliography,
superscriptaddress,
amsmath,
amssymb,
floatfix,
]{revtex4-1}
\usepackage[utf8]{inputenc}
\usepackage{microtype}
\usepackage{graphicx}
\usepackage{amsmath}
\usepackage{mathtools}
\usepackage[colorlinks=true,urlcolor=blue,citecolor=blue,linkcolor=blue]{hyperref}
\usepackage{natbib}
\usepackage{physics}
\usepackage{cleveref}
\usepackage{sidecap}
\usepackage[linesnumbered,lined,commentsnumbered]{algorithm2e}
\Crefname{algocf}{Algorithm}{Algorithms}
\usepackage{layouts}

\usepackage{xcolor}
\usepackage[normalem]{ulem}
\newcommand{\stkout}[1]{\ifmmode\text{\sout{\ensuremath{#1}}}\else\sout{#1}\fi}

\begin{document}

\title{Leakage in restless quantum gate calibration}
\date{\today}
\author{Conrad J. Haupt}
\affiliation{IBM Quantum, IBM Research Europe - Zurich, R\"uschlikon 8803, Switzerland}
\affiliation{ETH Z\"urich, 8093, Switzerland}
\author{Daniel J. Egger}
\email{deg@zurich.ibm.com}
\affiliation{IBM Quantum, IBM Research Europe - Zurich, R\"uschlikon 8803, Switzerland}

\begin{abstract}
    Quantum computers require high fidelity quantum gates.
    These gates are obtained by routine calibration tasks that eat into the availability of cloud-based devices.
    Restless circuit execution speeds-up characterization and calibration by foregoing qubit reset in between circuits. 
    Post-processing the measured data recovers the desired signal.
    However, since the qubits are not reset, leakage---typically present at the beginning of the calibration---may cause issues.
    Here, we develop a simulator of restless circuit execution based on a Markov Chain to study the effect of leakage.
    In the context of error amplifying single-qubit gates sequences, we show that restless calibration tolerates up to 0.5\% of leakage which is large compared to the $10^{-4}$ gate fidelity of modern single-qubit gates.
    Furthermore, we show that restless circuit execution with leaky gates reduces by 33\% the sensitivity of the ORBIT cost function developed by J. Kelly \emph{et al.} which is typically used in closed-loop optimal control~[Phys. Rev. Lett. \textbf{112}, 240504 (2014)].
    Our results are obtained with standard qubit state discrimination showing that restless circuit execution is resilient against misclassified non-computational states.
    In summary, the restless method is sufficiently robust against leakage in both standard and closed-loop optimal control gate calibration to provided accurate results.
\end{abstract}

\maketitle{}

\section{Introduction}

The performance of a quantum computer is benchmarked by its scale, quality, and speed.
These metrics are measurable by the number of qubits, the Quantum Volume~\cite{Cross2019, Jurcevic2021}, and the circuit layer operations per second~\cite{Wack2021}, respectively.
Crucially, the quantum gates that implement a quantum algorithm must be precisely calibrated to reach a high quality.
However, many quantum architectures, such as transmons~\cite{Koch2007}, embed a qubit in a large Hilbert space.
These extra states must be considered when calibrating gates~\cite{Motzoi2009, Schutjens2013}.
Fast gates speed-up quantum computations~\cite{Zhu2021, Weidenfeller2022} as long as they do not compromise quality, for example, by leaking qubit population out of the computational sub-space~\cite{Motzoi2009}.

On noisy quantum systems with weakly anharmonic qubits, leakage in single-qubit gates is avoided by DRAG pulses~\cite{Motzoi2009}.
However, even DRAG pulses do not fully mitigate leakage if they are too short.
Optimal control can further reduce the duration of single-qubit gates~\cite{Schutjens2013, Werninghaus2021, Zhu2021}.
In error correcting codes, leakage has recently become a focal point of research since it propagates to neighbouring qubits and degrades logical error rates~\cite{miao2022}.
E.g., Ref.~\cite{miao2022} shows how leakage spreads in the surface code~\cite{Fowler2012} and demonstrates that active leakage removal enables quantum error correction.
Leakage removal schemes for quantum error correcting codes have thus been designed, e.g., by emptying a frequency tunable qubit through a resonator~\cite{McEwen2021}, or by depopulating leakage states with an active reset~\cite{Battistel2021, Magnard2018, Egger2018, Geerlings2013}.
If leakage were not an issue, syndrome reset may not be needed as post-processing the measured outcomes accounts for the initial states~\cite{Wootton2022}.
Foregoing reset increases the error correction cycle rate. 
This may be necessary to demonstrate a practical advantage with quantum algorithms that offer a quadratic speed-up~\cite{Chakrabarti2021}.
Similarly, restless circuit execution for characterization and calibration does not reset the qubits in between circuit executions~\cite{Rol2017, Werninghaus2021b}. 
This enables closed-loop optimal control based on the measurement of sequences of Clifford gates~\cite{Kelly2014, Rol2017, Werninghaus2021}.
Indeed, the large amount of data needed is prohibitive if not gathered rapidly.
Restless circuit execution also speeds-up standard error amplifying gate calibration sequences, Randomized Benchmarking~\cite{Magesan2011, Magesan2012b} and Quantum Process Tomography~\cite{Tornow2022}. 

While leakage and its propagation in quantum error correction is the subject of intense research, little is known about its impact on restless characterisation and calibration.
Since restless foregoes reset one may thus wonder: ``How does leakage impact restless circuit execution?''.
Here, we therefore study the impact of leakage in restless calibration with a restless simulator that models measurement outcomes with a Markov Chain.
The simulator accounts for both unitary and non-unitary dynamics to capture processes such as $T_1$-decay that reduce leakage build-up.
We design single-qubit DRAG gates with a varying degree of leakage by changing the gate duration.
These gates are used in simulations of error-amplifying restless calibration experiments where we seek to determine if gate-errors are accurately measurable.
We observe that when leakage is too strong, first restless circuit execution fails to properly measure errors closely followed by standard circuit execution.
We also study the effect of leakage on restless closed-loop pulse shaping where we find that leakage reduces the sensitivity of the ORBIT cost function~\cite{Kelly2014} by a maximum of 1/3 compared to standard circuit execution with reset.

In Sec.~\ref{sec:restless_simulator} we present the restless circuit execution simulator.
In Sec.~\ref{sec:leakage_in_restless} we study the impact of leakage on error amplifying gate sequences measured with restless and standard circuit execution.
In Sec.~\ref{sec:rb} we investigate leakage in randomized-benchmarking based cost functions as used in optimal control~\cite{Rol2017, Werninghaus2021}.
We conclude in Sec.~\ref{sec:conclusion}.

\section{Restless simulator}\label{sec:restless_simulator}

As opposed to simulating quantum circuits that reset the qubits to $\ket{0}$ before each circuit, simulating a restless execution of quantum circuits requires that we (i) allow initial states other than $\ket{0}$ and (ii) account for and define the order in which the quantum circuits are executed.
Throughout this work, we assume that a list $[C_0,..., C_{K-1}]$ of quantum circuits is measured $N$ times. 
Each measurement of a circuit is called a shot.
Crucially, each circuit is measured once before the next round of shots is acquired, i.e., the $j^\text{th}$ shot for each circuit is executed before proceeding to the $j^\text{th}+1$ shot.
In restless, the readout of shot $j$ for quantum circuit $C_k$ is not followed by a reset of the qubits and the projected state serves as the initial state of the next circuit $C_{k+1}$.
The simulator must therefore produce time-ordered measurement outcomes $M_{kj}$, with $k$ and $j$ the circuit and shot indices, respectively.
We denote the time-ordered circuit execution number by $\zeta$ which is given by $\zeta=jK+k$ with $k=0,...,K-1$ and $j=0,...,N-1$.
The bit-wise exclusive OR of two consecutive outcomes, i.e. bit strings, indicates which qubits underwent a state change due to the execution of a quantum circuit.
For example, if two consecutive bitstring are \texttt{`0101'} and \texttt{`1100'} then only qubits 0 and 3 changed state~\footnote{We use a little-endian qubit ordering.}.
The probability of observing a state change is sufficient to perform many characterization and calibration tasks as discussed in Ref.~\cite{Tornow2022}.

We now present a methodology to simulate restless circuit execution capable of including non-computational states to model leakage.
We denote the basis of quantum states by $\mathcal{B}^{n_q}$, for example, $\mathcal{B}=\{\ket{0}, \ket{1}, \ket{2}\}$ to simulate transmons modeled by three levels.
$n_q$ is the number of transmons.
Our restless simulator samples from a Markov Chain to produce a list of measurement outcomes $M_{kj}\in\mathcal{M}^{n_q}$, for each shot $j=0,...,N-1$ and circuit $C_k$ with $k=0,...,K-1$.
Here, $\mathcal{M}$ is the set of possible outcomes such as $\{\texttt{`0'}, \texttt{`1'}\}$ for qubits or $\{\texttt{`0'}, \texttt{`1'}, \texttt{`2'}\}$ if qutrit discrimination is enabled.

\begin{SCfigure*}
\centering
\includegraphics[width=1.5\columnwidth]{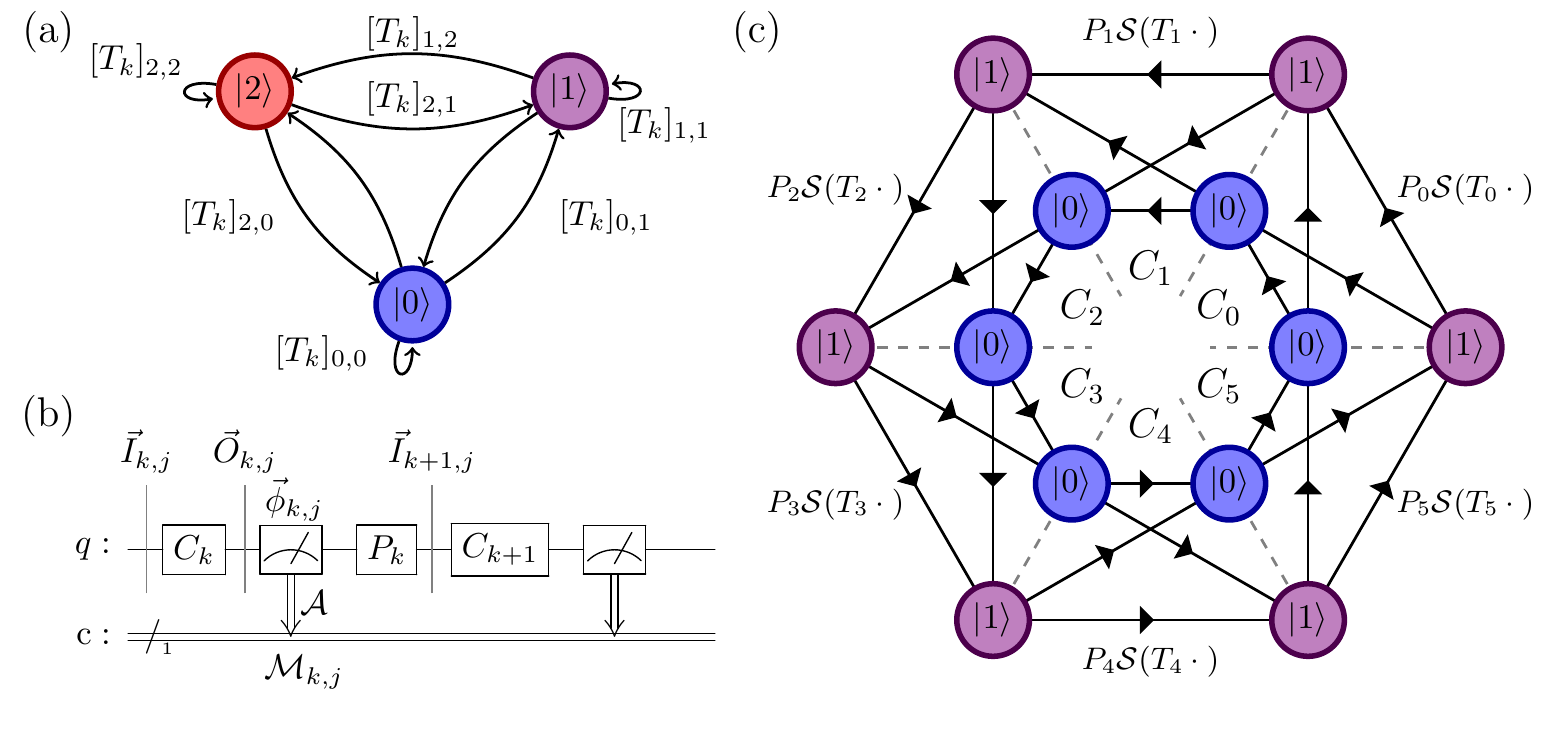}
\caption{
Restless simulator.
(a) Depiction of the transition matrix $T_k$ with a basis $\mathcal{B}=\{\ket{0}, \ket{1}, \ket{2}\}$.
The elements of $T_k$ and $\mathcal{B}$ are shown as edges and nodes, respectively.
(b) Circuit to illustrate the notation describing the restless simulator. 
(c) Depiction of the cyclic Markov Chain of a restless simulation of six circuits $C_0$, ..., $C_5$.
The arrows in each sector correspond to the transition function $\mathcal{S}[P_k\mathcal{S}(T_k\cdot)]$ which is applied to $\vec{I}_{k,j}$, for example, $\mathcal{S}[P_2\mathcal{S}(T_2\vec{I}_{2j})]$.
Each sector corresponds to a circuit $C_k$.
\label{fig:simulator}
}
\end{SCfigure*}

Before drawing shots, the restless simulator first computes a transition matrix $T_k$ for each quantum circuit $C_k$. 
Since we assume that $C_k$ is followed by a strong measurement we ignore any coherence in the post-measurement states.
The transition matrix element $ [T_k]_{\mu\nu}$ is thus the probability that the measurement will project the quantum state into state $\ket{\mu}\in\mathcal{B}$ given the input basis state $\ket{\nu}\in\mathcal{B}$, i.e., the entries of $T_k$ are
\begin{equation}\label{eq:trans_mat_entries}
    [T_k]_{\mu\nu}={\rm Tr}\{\ket{\mu}\!\!\bra{\mu}C_k(\ket{\nu}\!\!\bra{\nu})\},
\end{equation}
see Fig.~\ref{fig:simulator}(a).
Here, it is understood that $C_k$ is a completely positive trace preserving map of the $k^\text{th}$ circuit.

Next, to draw time-ordered shots we describe the input state to the $j^\text{th}$ shot of the $k^\text{th}$ circuit as an input vector $\vec{I}_{kj}$ where all entries are zero other than the one corresponding to the input state $\ket{\nu}$, i.e., $[\vec{I}_{kj}]_\nu=1$.
We assume that the very first input state is the ground state, i.e., $[\vec{I}_{00}]_0=1$.
The probability $p_\mu$ to measure state $\ket{\mu}\in\mathcal{B}$ for circuit $C_k$ is stored in a vector $\vec{O}_{kj}$ with entries $[\vec{O}_{kj}]_\mu=p_\mu$ that satisfy $\sum_\mu p_\mu=1$.
The two probability vectors $\vec{I}_{kj}$ and $\vec{O}_{kj}$ are related by the transition matrix $\vec{O}_{kj}=T_k\vec{I}_{kj}$.

To create a measurement outcome for shot $j$ of circuit $k$ we (i) sample a basis state from $\mathcal{B}$, now corresponding to and labeled by $\vec{\phi}_{kj}$, according to the probabilities in $\vec{O}_{kj}$ and (ii) apply a basis state to shot labelling function.
Step (i) captures the collapse of the wave-function that occurs during a strong measurement.
Measurement assignment errors are captured by step (ii) in which we multiply $\vec{\phi}_{kj}$ by a $|\mathcal{M}|\times|\mathcal{B}|$ dimensional matrix $\mathcal{A}$. 
A shot $M_{kj}\in\mathcal{M}$ is created by sampling from the probabilities $\mathcal{A}\vec{\phi}_{kj}$. 

Finally, to create the input state to the next circuit $C_{k+1}$ we add post-measurement errors, such as relaxation~\cite{Picot2008} or measurement-induced state transitions~\cite{Slichter2012,Hatridge2013}, modelled by an additional transition matrix $P_{k}$.
The input basis state to circuit $C_{k+1}$ is thus described by $\vec{I}_{k+1,j}=\mathcal{S}(P_k\vec{\phi}_{kj})$.
Here, $\mathcal{S}$ denotes a sampling that converts a vector of probabilities to a vector that corresponds to a basis state.
The steps of the restless simulator are summarized in Algorithm~\ref{alg:restless_sim} and illustrated in Fig.~\ref{fig:simulator}(b).
Note that the size of the transition matrices scale as $2^{n_q}$.
This is acceptable since the simulator is designed to study characterization and calibration experiments which run on a small number of transmons.

\begin{algorithm}
\caption{Restless simulation of $K$ circuits on a single transmon modelled as a qutrit.
Note that we do not store the intermediate variables $\vec{I}_{kj}$, $\vec{O}_{kj}$, and $\vec{\phi}_{kj}$ and simply reuse the same memory location $\vec{I}$, $\vec{O}$, and $\vec{\phi}$, respectively.
}\label{alg:restless_sim}
\SetKwInOut{Input}{input}\SetKwInOut{Output}{output}
\SetKwFunction{ComputeTransitionMat}{compute\_transition\_matrices}
\Input{List of $K$ circuits $[C_0,C_1,\ldots{},C_{K-1}]$,\\maximum number of shots $N$,\\measurement-assignment matrix $\mathcal{A}$,\\and post-measurement transition-matrices $P_k$.}
\Output{List of measurement outcomes $M_{kj}$.}
\BlankLine
\emph{Compute transition matrices and store.}

\For{$0\leq{}k<K$}{
$T_k\gets $\ComputeTransitionMat{$C_k$}
}
\BlankLine
\emph{Start in the ground-state.}

$\vec{I}\gets {(1, 0, 0)}^T$

\For{each shot $0\leq{}j<N$}{
    \For{each circuit index $0\leq{}k<K$}{
        $\vec{O}\gets{}T_k\vec{I}$
        
        $\vec{\phi}\gets{}\mathcal{S}(\vec{O})$
        
        $M_{kj}\gets{}\mathcal{S}(\mathcal{A}\vec{\phi})$
        
        $\vec{I}\gets{}\mathcal{S}(P_k\vec{\phi})$
    }
}
\Return Measurement outcomes $M_{kj}$ grouped by $k$ and sorted by $j$.
\end{algorithm}

To clarify the notation, depicted in Fig.~\ref{fig:simulator}(b), we now provide a simple qubit-based example, i.e., $\mathcal{B}=\{\ket{0}, \ket{1}\}$ and $\mathcal{M}=\{\texttt{`0'},\texttt{`1'}\}$.
Suppose that circuit $C_k$ applies an ideal Hadamard gate.
Its transition matrix is thus $[T_k]_{\mu\nu}=1/2$ with $\mu,\nu\in\{0,1\}$.
If the input state is $\ket{1}$ for shot $j$, i.e., $\vec{I}_{kj}=(0, 1)^T$, then the output vector is $\vec{O}_{kj}=(1/2, 1/2)^T$.
There is therefore a 50\% probability of sampling either $\ket{0}$ or $\ket{1}$ after a strong measurement.
Furthermore, if this sampling yields state $\ket{1}$ and the readout assignment is perfect, i.e. $\mathcal{A}$ is the $2\times2$ identity matrix, then the measurement outcome is \texttt{`1'}.
Without post-measurement errors, the input to the next circuit is state $\ket{1}$ described by $\vec{I}_{k+1,j}=(0, 1)^T$.
As example, the Markov Chain for six circuits running on a transmon modelled as a qubit is depicted in Fig.~\ref{fig:simulator}(c). 

\section{Leakage in restless Calibration\label{sec:leakage_in_restless}}

Since qubits are not reset to the ground-state during restless execution leakage may impact performance.
For instance, when calibrating a gate with a low level of leakage transitions to $\ket{2}$ occur infrequently but if a transition does occur then the probability to return to the computational subspace is also low.
State relaxation induced by $T_1$ mitigates the impact of such leakage events. 
However, calibration circuits typically have few gates each lasting of the order of $10$ -- $100~{\rm ns}$.
The duration of the circuits we are interested in are therefore orders of magnitude shorter than current $T_1$ times which are well in excess of $100~\mu{\rm s}$~\cite{Fischer2022}.

We now simulate calibration experiments for both standard and restless circuit execution with the restless simulator described in Sec.~\ref{sec:restless_simulator}.
To model standard circuit execution including the reset, we chose a post measurement matrix
\begin{align}
P_k=\begin{pmatrix}
    1 & 1 & 1 \\
    0 & 0 & 0 \\
    0 & 0 & 0
\end{pmatrix}
\end{align}
that always results in the ground state.
This assumes an ideal reset and a three-level model of the transmon.
$P_k$ is the identity matrix for ideal restless circuit execution, i.e., without decoherence and measurement-induced state transitions.

\subsection{System and setup\label{sec:setup}}

\begin{figure}[tb]
    \centering
    \includegraphics[width=0.99\columnwidth,clip,trim=7 5 10 5]{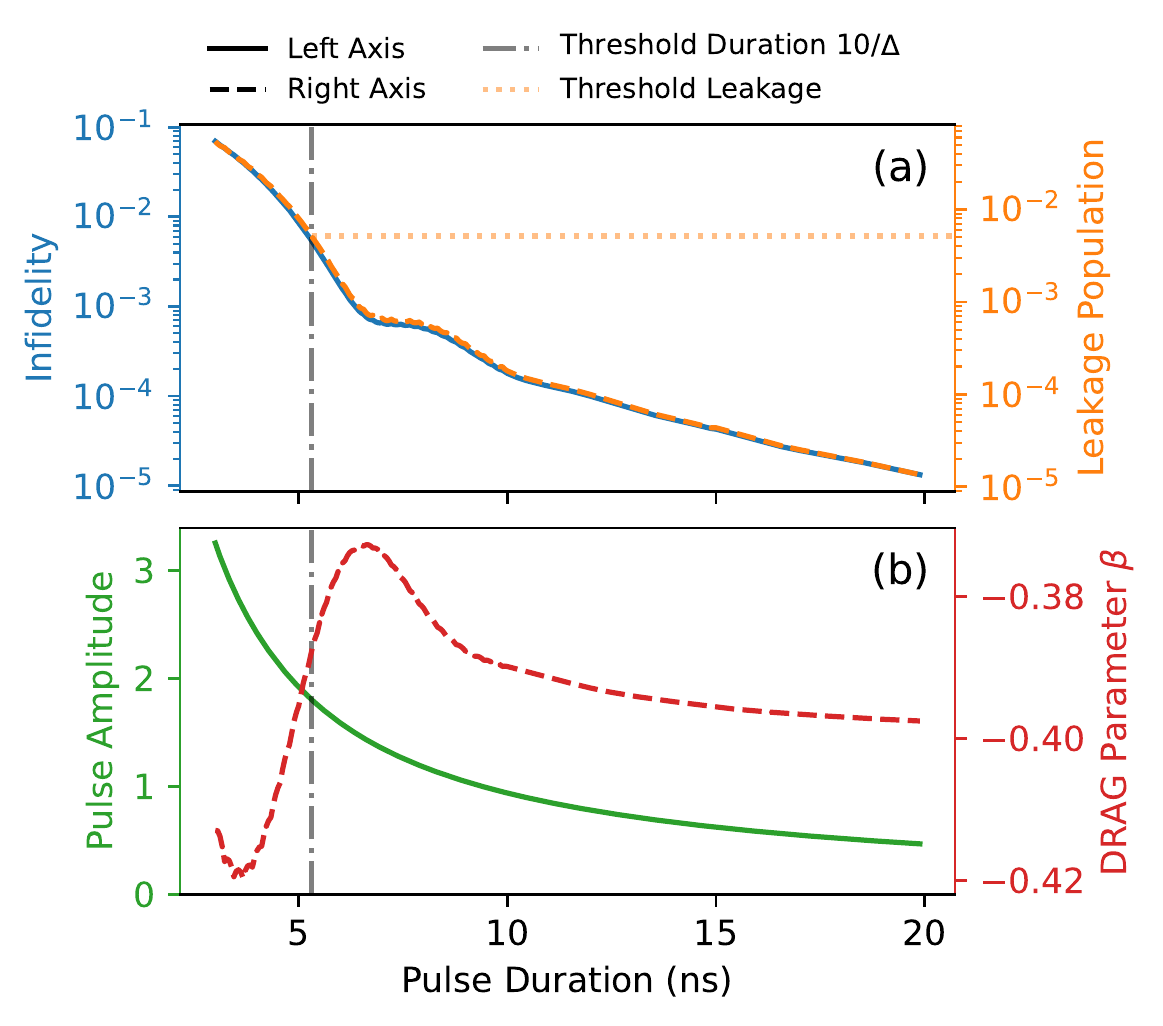}
    \caption{
    Simulated performance of DRAG pulses with different durations $\tau$.
    The amplitude and DRAG parameter are found by minimizing the infidelity $1 - \Phi$ of the resulting unitary in the qubit subspace. 
    (a) The infidelity of $U_\tau$ and population of the $\ket{2}$ state, i.e., $|\langle2|U_\tau|0\rangle|^2$. 
    (b) The optimized pulse amplitudes and DRAG parameters $\beta$.}
    \label{fig:optimized_pulse_performance}
\end{figure}

We study a fixed-frequency transmon with Hamiltonian
\begin{align}\label{eqn:hamiltonian}
\hat H = \omega\hat{a}^\dagger\hat{a} + \frac{\Delta}{2}\hat{a}^\dagger\hat{a}^\dagger\hat{a}\hat{a} +  \lambda\Omega(t)(\hat{a}^\dagger+\hat{a})
\end{align}
and retain the first three levels, i.e., $\mathcal{B}=\{\ket{0},\ket{1},\ket{2}\}$.
Here, $\omega$ is the transition frequency between the $\ket{0}$ and $\ket{1}$ states of the transmon.
$\hat a^\dagger$ and $\hat a$ are the creation and annihilation operators, respectively.
The anharmonicity $\Delta$ is $-300~{\rm MHz}$ and the control-line-qubit coupling rate $\lambda$ is $100~{\rm MHz}$.
$\Omega(t)$ is the dimensionless control pulse.
We design single-qubit DRAG pulses~\cite{Motzoi2009} of different durations $\tau$ to implement $X$ rotations with a varying amount of leakage.
The in-phase component $\Omega_x(t)$ of $\Omega(t)$ is a Gaussian function with standard deviation $\sigma$.
The quadrature is the derivative of $\Omega_x$ scaled by the DRAG parameter $\beta$.
The ratio $\tau/\sigma$ is fixed at $4$. 
For each pulse, we compute the time-evolution operator $U_\tau$ in the qutrit space $\mathcal{B}$ by computing the time-ordered integral of Eq.~(\ref{eqn:hamiltonian}) in the rotating frame of the qubit with Qiskit Dynamics~\cite{QiskitDynamics}.
We calibrate the amplitude of $\Omega_x$ and the DRAG parameter~$\beta$ by maximizing the process fidelity 
\begin{align}\label{eqn:f_qpt}
\Phi=\frac{1}{4}|{\rm Tr}\{\mathbb{P}U_\tau^\dagger{}\mathbb{P}^\dagger U_\text{target}\}|^2.
\end{align}
Here, $\mathbb{P}$ is the projector onto the computational subspace $\{\ket{0},\ket{1}\}$.
As target unitary $U_\text{target}$ we chose a $\pi$-rotation around the $x$-axis labelled by $R_x(\pi)$ or $X$.
The optimized DRAG pulses have monotonically decreasing leakage, measured as $|\langle2|U_\tau|0\rangle|^2$, ranging from $5.46\cdot 10^{-2}$ at $\tau=3~{\rm ns}$ to $1.35\cdot 10^{-5}$ at $\tau=20~{\rm ns}$, see Fig.~\ref{fig:optimized_pulse_performance}.
This allows us to vary the amount of leakage in the experiments we study by changing the pulse duration $\tau$.

We assume that the readout of the quantum computer is setup to discriminate qubit states. Therefore, the discriminator in the readout chain of the transmon erroneously classifies the $\ket{2}$ state as a \texttt{`1'} outcome and thus $\mathcal{M}=\{\texttt{`0'}, \texttt{`1'}\}$.
The assignment matrix is thus
\begin{equation}\label{eqn:meas_matrix}
    \mathcal{A} = \begin{pmatrix}
        1&0&0\\
        0&1&1\\
    \end{pmatrix}.
\end{equation}
This corresponds to the default operation of a superconducting qubit processor.
The measurement process that results in Eq.~(\ref{eqn:meas_matrix}) is discussed in Appendix~\ref{app:discrimination}.

\subsection{Amplitude calibration}\label{sec:amp_cal}

\begin{figure*}[thb!]
    \centering
    \includegraphics[width=\linewidth,clip,trim=0 4.8 0 0]{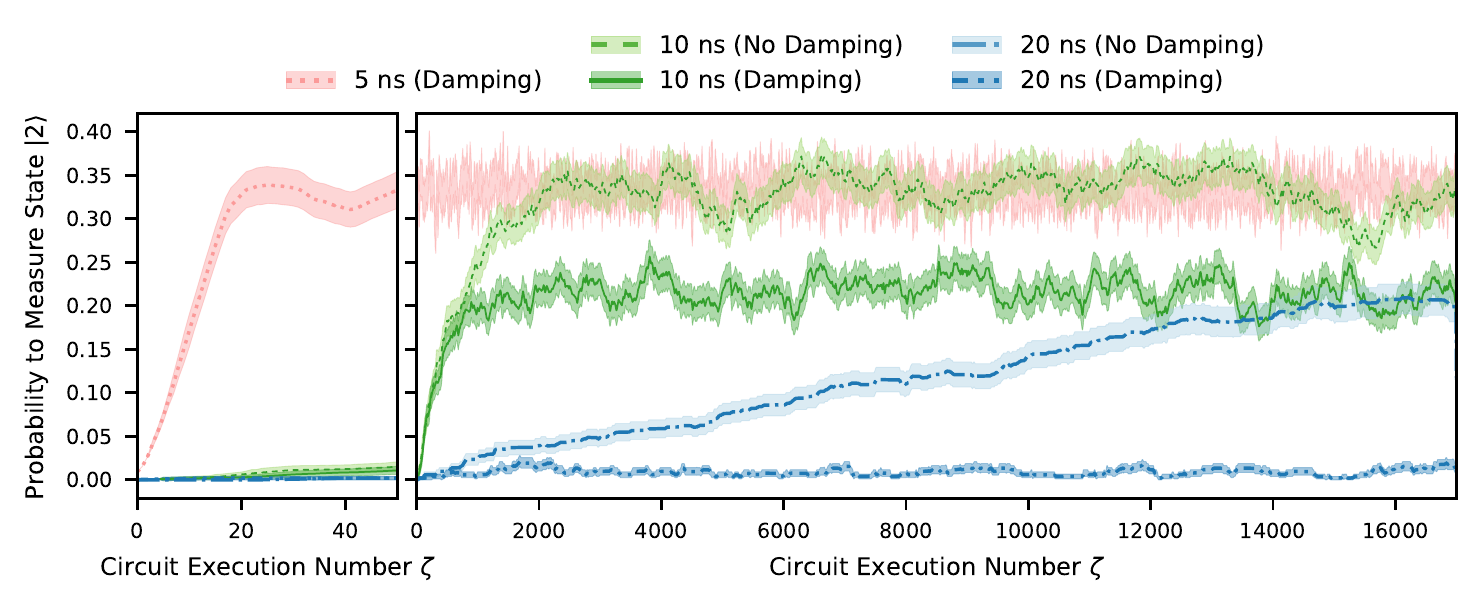}
    \caption{Probability to measure the $\lvert{}2\rangle{}$ state over time, as determined by $512$ realizations of a restless error-amplification sequence of 17 individual circuits. 
    The probabilities are taken as the moving average of the qutrit measurement outcomes over all realizations with a window size of $16$ in $\zeta$.
    The standard-deviation of the mean is shown as shaded areas.
    For the shortest pulse duration in the right-axis, only the standard deviation is plotted, as its probabilities are very noisy and diminish the readability of the figure.
    The results for the $5~{\rm ns}$ pulse without damping are similar to the results with damping and are not shown for readability purposes.
    Though only 20 circuits have been executed by $\zeta=20$, per realization, this includes $146$ leaky gates which all contribute to the accumulated leakage which is higher than indicated by the single-gate leakage level in Fig.~\ref{fig:optimized_pulse_performance}, see explanation in Appendix~\ref{app_tmats}.
    }
    \label{fig:fineamplitude_timedomain_leakage}
\end{figure*}

We study leakage in an amplitude-error amplifying gate sequence done with standard and restless circuit execution.
The amplitude of the $X$ gate is calibrated by the gate sequence $\sqrt{X}$ -- $[X]^n$ followed by a measurement for varying $n$.
This creates states that ideally lie on the equator of the Bloch sphere thereby maximizing the sensitivity to small rotation errors when measured in the $Z$ basis.

\subsubsection{Leakage build-up in restless execution\label{sec:leakage_buildup}}

To study leakage, we consider the amplitude-error amplifying experiment with 1000 shots and $K=17$ circuits.
We compute the time-ordered probabilities \smash{$p^{(2)}_{\zeta}$} with which each shot $j=0,\ldots,999$ and circuit $k=0,\ldots,16$ results in \smash{$\vec{\phi}_{\zeta}=(0,0,1)^T$} corresponding to a $\ket{2}$ state.
$p_\zeta^{(2)}$ is given as a function of the circuit execution number $\zeta$ given by $jK+k$.
Since the restless simulator is a Markov Chain the probability \smash{$p^{(2)}_{\zeta}$} depends only on the previous outcome.
Therefore, to obtain the behaviour of \smash{$p^{(2)}_{\zeta}$} we simulate 512 realizations of the experiment, and for each shot and circuit we estimate \smash{$p^{(2)}_{\zeta}$} as $n_2/512$ with $n_2$ the number of times circuit execution $\zeta$ produced state $\ket{2}$.
The simulation is repeated for pulses with a duration of $5~\rm{ns}$, $10~\rm{ns}$, and $20~\rm{ns}$ to change the amount of leakage in accordance with Fig.~\ref{fig:optimized_pulse_performance}.
We observe a build-up of leakage over the course of a restless experiment.
\smash{$p^{(2)}_{\zeta}$} increases over time for all pulse durations and fluctuates around a fixed average for short pulses and high circuit execution numbers, see Fig.~\ref{fig:fineamplitude_timedomain_leakage}.
The $5~\rm{ns}$ and $10~\rm{ns}$ pulses both oscillate at \smash{$p^{(2)}=1/3$}.
Though the population of the $\ket{2}$ state for the $20~\rm{ns}$ pulse --- which has a leakage of $10^{-5}$ --- does not settle within 1000 circuit executions, the probability \smash{$p^{(2)}$} is approximately $20\%$ towards the last measured shots.

In hardware, energy dissipates from the transmons relaxing the quantum states towards $\ket{0}$ thereby suppressing leakage.
To account for this leakage suppression, we repeat the simulation with an amplitude damping channel, as described in Ref.~\cite{chessa2021}, with relaxation times between the qutrit states of $100~\rm{\mu{}s}$ and $71~\rm{\mu{}s}$ for the $1\rightarrow{}0$ and $2\rightarrow{}1$ transitions, respectively.
The rate of the $2\rightarrow{}0$ transition is set to 0 in accordance with experimental observations~\cite{Fischer2022}.
Appendix~\ref{app:damping} describes how the amplitude damping channel is included in the restless simulator.
The amplitude damping channel reduces the $\ket{2}$ state population, and even suppresses it for the longest pulse duration, see Fig.~\ref{fig:fineamplitude_timedomain_leakage}.
Though amplitude damping does not fully suppress the $\ket{2}$ state population for the shorter pulse durations, it does reduce the level at which \smash{$p^{(2)}$} settles, compare the damping and no damping lines in Fig.~\ref{fig:fineamplitude_timedomain_leakage}.
The amplitude damping channel reduces the fixed average for the $10~\rm{ns}$ pulse to $21.7\%$.
This reduction with an amplitude damping channel is not observed for $5~\rm{ns}$ pulses.
The amplitude damping channel effectively suppresses \smash{$p^{(2)}$} for $20~{\rm ns}$ pulses.

\subsubsection{Calibration}

We now investigate if the observed leakage build-up prevents accurately measuring rotation errors $\delta\theta$ and calibrating the pulse amplitude.
We create a set of $U_\tau$ rotations, as described in Sec.~\ref{sec:setup}, with $U_\text{target}=R_x(\pi[1+\varepsilon])$ to intentionally introduce a rotation error~$\varepsilon$.
We measure $\varepsilon$ with the error amplifying gate sequence under varying degrees of leakage controlled by changing $\tau$ from $3~{\rm ns}$ to $20~{\rm ns}$, in accordance with Fig.~\ref{fig:optimized_pulse_performance}.
The restless measured shots, with outcome \texttt{`2'} erroneously classified as \texttt{`1'}, are post-processed by taking the exclusive OR between two consecutive outcomes~\cite{Tornow2022}.
The resulting signal, in the absence of leakage, gives the probability that a circuit changed the state of a qubit.
A fit of the signal to the function 
\begin{align}\label{eqn:error_amp_fit}
a\cos\left[(\theta_{t} + \delta\theta)n - \frac{\pi}{2}\right] + b
\end{align}
reveals the rotation error $\delta\theta$ which we compare to $\varepsilon$.
Here, $a$ and  $b$ are fit parameters and $\theta_t=\pi$ is the target rotation angle per gate.

\begin{figure}[htbp!]
    \centering
    \includegraphics[width=\linewidth,clip,trim=12 0 10 0]{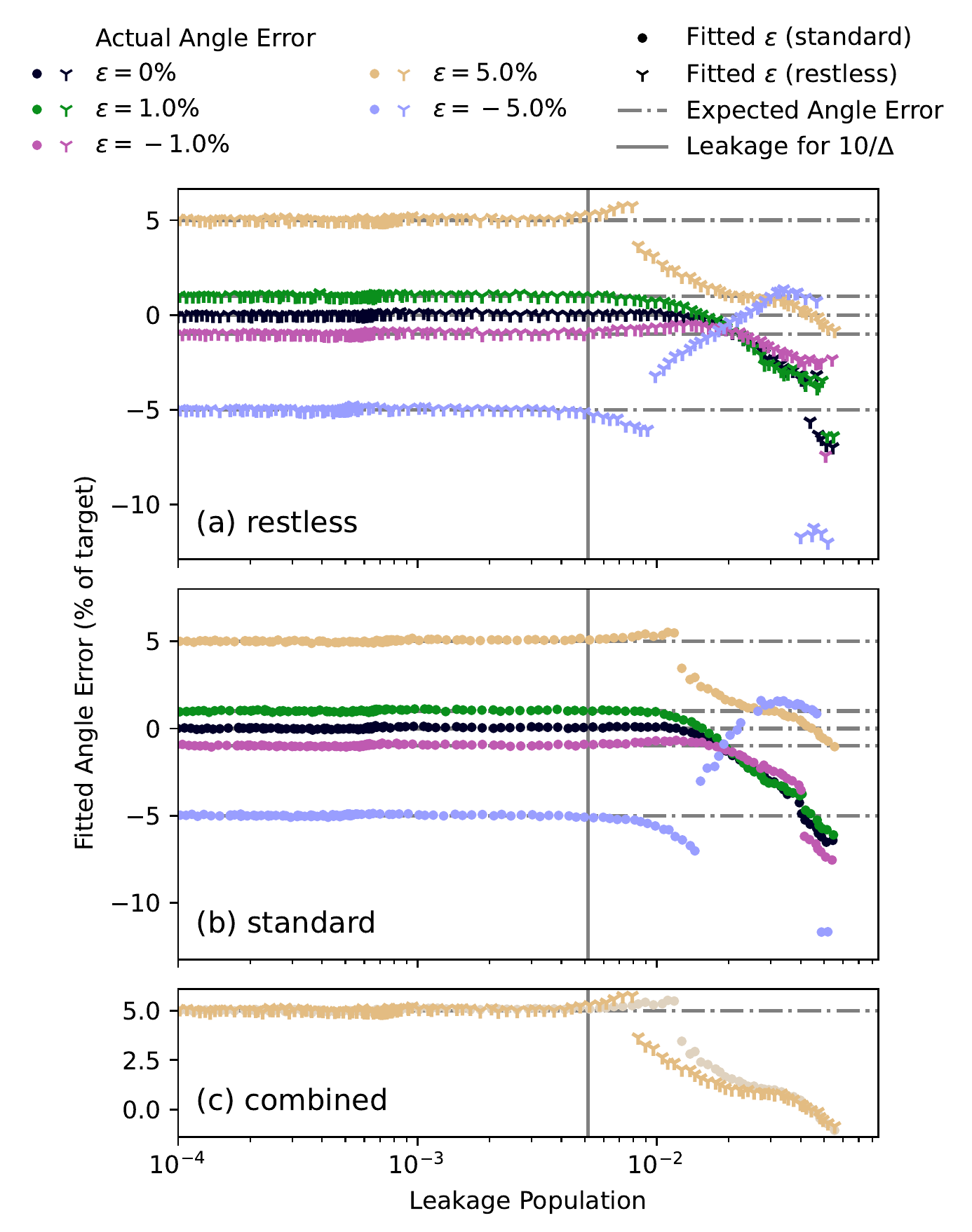}
    \caption{Measured angle-error $\delta\theta$ as a function of leakage.
    The leakage strength is varied by changing the duration $\tau$ of the pulse implementing $R_x(\pi[1+\varepsilon])$.
    The error amplification sequence is run using restless (a) and standard circuit execution (b).
    The vertical line indicates the leakage of a pulse with a duration of $10/\Delta$.
    The bottom panel combines restless and leakage results at $\varepsilon=5\%$ rotation error, where the standard execution method markers are made slightly lighter to improve contrast.
    }
    \label{fig:fineamplitude_vs_leakage}
\end{figure}

Below a duration threshold of $10/\Delta$, i.e., for leakage greater than 0.5\%, discrepancies between the measured angle error $\delta\theta$ and the actual error $\varepsilon$ emerge, see Fig.~\ref{fig:fineamplitude_vs_leakage}(a).
The $10/\Delta$ duration threshold is also used as an indicator in prior work as a point below which DRAG pulses suffer from significant leakage~\cite{Werninghaus2021, Motzoi2009}.
When run with standard execution, discrepancies between $\delta\theta$ and $\varepsilon$ are also observed but for pulses with more then 1\% leakage, see Fig.~\ref{fig:fineamplitude_vs_leakage}(b). 
This indicates that restless execution makes the error-amplification sequence only a little more sensitive to leakage than standard execution, see Fig.~\ref{fig:fineamplitude_vs_leakage}(c).
For the example considered here, restless and standard measurements can tolerate 0.5\% and 1\% leakage, respectively.
In both cases this is a large amount of leakage since current superconducting systems achieve single-qubit gate errors of $10^{-4}$ after calibration.

In practice, error amplifying calibration sequences are done iteratively. 
At each iteration, $\delta\theta$ is measured and the pulse amplitude is updated by multiplying it by $\theta_t/(\theta_t +\delta\theta)$.
The calibration stops when $\delta\theta$ falls below a set threshold.
We numerically investigate this convergence with rotation errors $\varepsilon\in\{0\%, 1\%, 5\%\}$ and pulses that last $3~{\rm ns}$, $3.5~{\rm ns}$, $4.5~{\rm ns}$, and $10~{\rm ns}$ corresponding to a leakage of $5.46\%$, $3.71\%$, $1.47\%$, and $0.0179\%$, respectively.
At each iteration we calculate the infidelity of the pulse $E=1-\Phi$ and compare it to the infidelity $E_\text{opt}$ of the best possible pulse designed with $\varepsilon=0\%$ rotation error.

\begin{figure}
    \centering
    \includegraphics[width=\linewidth]{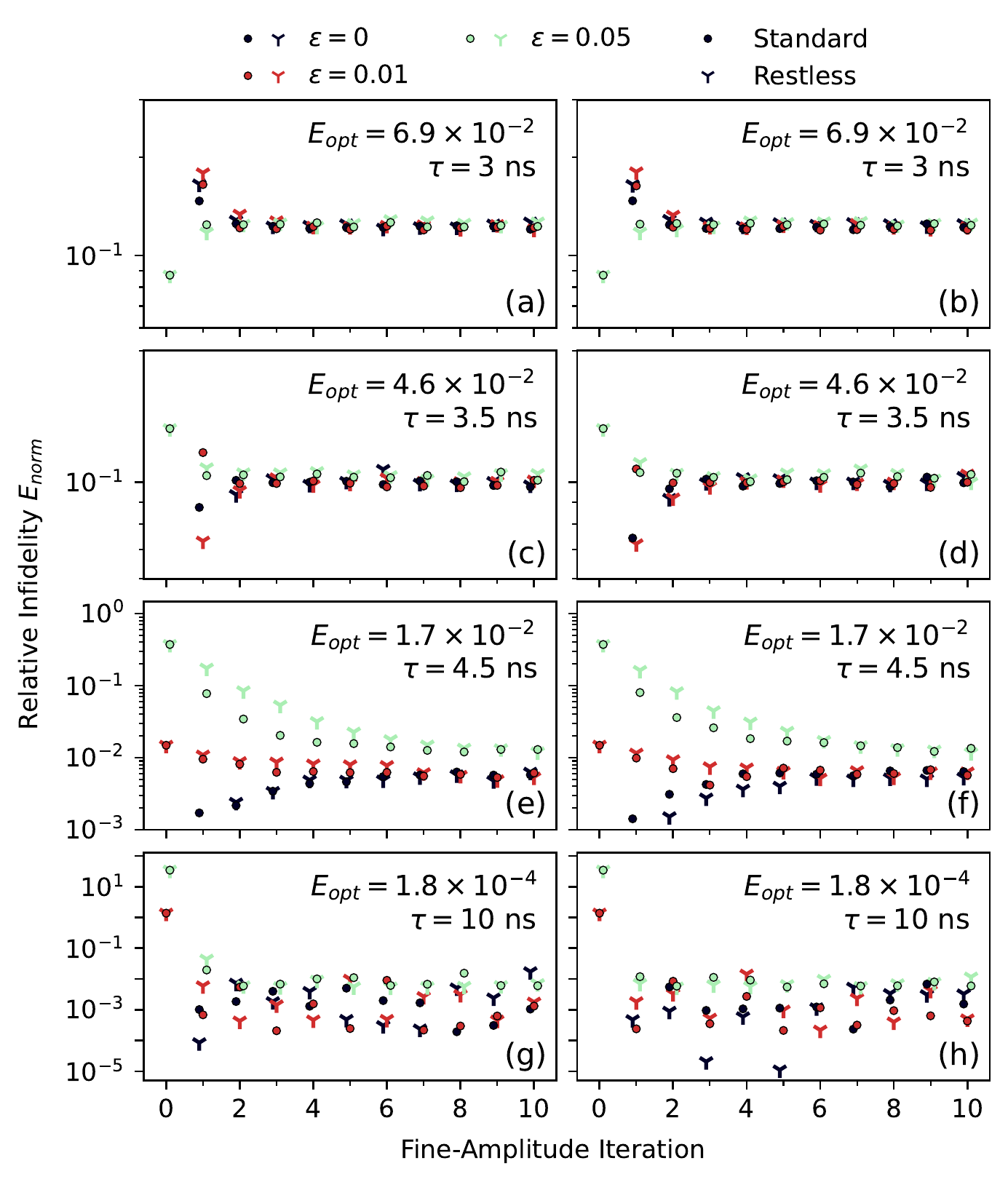}
    \caption{Normalized relative infidelity $E_\text{norm}$ for iterative fine-amplitude experiments for different pulse durations $\tau$. 
    The computed infidelities $E=1-\Phi$ are normalized with respect to the optimal infidelity $E_\text{opt}$ obtained by optimizing $\Phi$ at $\varepsilon=0$, i.e., $E_\text{norm}=(E - E_\text{opt})/E_\text{opt}$. 
    The optimal infidelity $E_\text{opt}$ corresponds to the zeroth-iteration infidelity for $\varepsilon=0$. 
    Simulations were carried out for each pulse duration without (a, c, e, g) and with (b, d, f, h) an amplitude damping channel; as well as with and without restless circuit execution.
    \label{fig:iterative_amp}}
\end{figure}

For all levels of leakage, the infidelity of the pulses converge to a higher level than that of the pulse designed with $\varepsilon=0$.
I.e., all data points in Fig.~\ref{fig:iterative_amp}, except at iteration zero with $\varepsilon=0$, have a finite value.
Unsurprisingly, this implies that an imperfect calibration experiment makes an ideal pulse worse.
For example, the infidelity of $4.5~{\rm ns}$ pulses designed with $\varepsilon=0$ increases with iteration number while the infidelity of pulses with $\varepsilon=5\%$ decreases, compare the black and light green (light gray) markers in Fig.~\ref{fig:iterative_amp}(e) and (f).
As the pulse-duration increases and leakage decreases the re-calibrated pulses are closer to those designed without an error, i.e., $\varepsilon=0$, compare the infidelity of the final iteration in Fig.~\ref{fig:iterative_amp}(a), (c), (e), and (g).
Interestingly, there is little difference between standard and restless circuit execution, compare round and cross markers in Fig.~\ref{fig:iterative_amp}, with restless execution requiring a few more iterations to converge in some cases, e.g., Fig.~\ref{fig:iterative_amp}(e).
The presence of an amplitude damping channel does not impact these results, compare right and left panels in Fig.~\ref{fig:iterative_amp}.
The relative infidelities plotted in Fig.~\ref{fig:iterative_amp} are normalized to the optimal infidelity $E_\text{opt}$. 
This makes the variance of the $10~\mathrm{ns}$ pulses appear larger since their optimal infidelity is $1.8\cdot 10^{-4}$, i.e., two orders of magnitude lower than the shorter pulses.
This fine amplitude calibration example shows that in the presence of leakage restless circuit execution can still calibrate pulses.

\subsection{ORBIT Gate Calibration\label{sec:rb}}

In Sec.~\ref{sec:leakage_in_restless} we studied error amplifying gate sequences which typically have a small number of gates.
To explore sequences with many gates, we turn to a quantum optimal control use case.
A typical application of quantum optimal control is pulse shaping to implement a target gate~\cite{Khaneja2005, Egger2014, Glaser2015, Chasseur2015, Boscain2021, Carvalho2021}.
Parameter drift and model inaccuracies in superconducting qubit systems render open-loop optimal control inaccurate in practice~\cite{Egger2014b}.
One must therefore improve model identification~\cite{Wittler2021} and use closed-loop optimisation directly on the hardware~\cite{Egger2014b, Kelly2014}.
Evaluating a fidelity with randomized benchmarking (RB)~\cite{Magesan2011} or quantum process tomography requires~\cite{Banaszek1999} a large number of circuit executions.
Consider, for example, RB. 
The Error per Clifford $r_c$ for a given RB sequence is obtained by fitting
\begin{align}\label{eq:rb_ideal}
    \mathcal{F}_\text{seq}(m)=A\alpha^m+B
\end{align}
to a qubit population measured after random sequences of Clifford gates of variable length $m$.
The ideal sequences compose to the identity.
Here, $A$ and $B$ absorb state preparation and measurement (SPAM) errors and the Error per Clifford $r_c$ is linearly related to the depolarizing parameter $\alpha$, e.g., $\alpha=1-2r_c$ for single-qubit gates.
Optimized Randomized Benchmarking for Immediate Tune-up (ORBIT) recognizes that $\mathcal{F}_\text{seq}$ monotonically decreases with increasing $r_c$.
ORBIT thus calibrates gates by evaluating multiple sequences of Clifford gates with the same fixed depth~\cite{Kelly2014}.
I.e., $\mathcal{F}_\text{seq}(m)$ at fixed $m$ is a hardware efficient cost function for closed-loop optimal control.
ORBIT evaluates changes in gate fidelities faster than RB; especially when combined with restless circuit execution~\cite{Rol2017, Werninghaus2021}.
ORBIT can optimize single- and two-qubit gates on superconducting transmon quantum hardware~\cite{Kelly2014, Rol2017, Werninghaus2021}.
In Ref.~\cite{Werninghaus2021} a short high-leakage DRAG pulse initializes a closed-loop ORBIT calibration in which pulse samples are further optimized to reduce leakage.
We therefore study the effect of leakage on restless ORBIT and its sensitivity to variations in gate fidelities.

In our simulation, each Clifford gate in the ORBIT sequence is built from the four gates $\{R_x(\pm \pi/2)$, $R_y(\pm \pi/2)\}$.
These gates have non-zero infidelity and leakage owing to limitations of the DRAG pulse shape and the finite pulse duration.
We engineer leakage by varying the pulse durations and DRAG parameter.
For each duration in $\{3, 3.5, 4, 4.5, 5, 10, 20\}~{\rm ns}$ we numerically optimize the process fidelity $\Phi$ of Eq.~(\ref{eqn:f_qpt}).
Next, by scaling the resulting optimal DRAG parameter $\beta_{opt}$ for each pulse by $30$ prefactors equidistant in $[-2, 2]$, we further create variations in $\Phi$ that restless ORBIT should be sensitive to.
We measure $\mathcal{F}_\text{seq}(m)$ by averaging $100$ random single-qubit Clifford gate sequences of depth $m+1$ with $m\in\{30,60,90,120\}$ that ideally compose to the identity~\cite{Mckay2019}.
We sample $1000$ shots from the resulting circuits with the restless simulator, with and without restless circuit execution, and with and without a damping channel as described in Appendix~\ref{app:damping}.
Next, we compare the sequence fidelity $\mathcal{F}_\text{seq}$ to the average error per Clifford gate $r_c$ which we compute from the process fidelity $\Phi$, averaged over the four rotations $R_x(\pm \pi/2)$, $R_y(\pm \pi/2)$.
The average error per Clifford is related to the average fidelity of the rotations by $r_c\simeq1 - \mathcal{F}_\text{avg}(\Phi)^{N_c}$.
Here, $N_c\approx2.1666$ is the average number of $R_{x,y}$ rotations per Clifford gate in our sequences of Clifford gates. 
The average gate fidelity $\mathcal{F}_\text{avg}$ is related to the process fidelity $\Phi$ by $\mathcal{F}_\text{avg}=(d\Phi+1)/(d+1)=(2\Phi+1)/3$ where we assume $d=2$ for single-qubit gates~\cite{Horodecki1999, Magesan2011b}. 
Therefore, we compare $\mathcal{F}_\text{seq}$ to $r_c(\Phi)=1-[(2\Phi+1)/3]^{N_c}$ for various pulse durations and DRAG parameters.

\begin{figure}
    \centering
    \includegraphics[width=\linewidth, clip,trim=7 0 10 0]{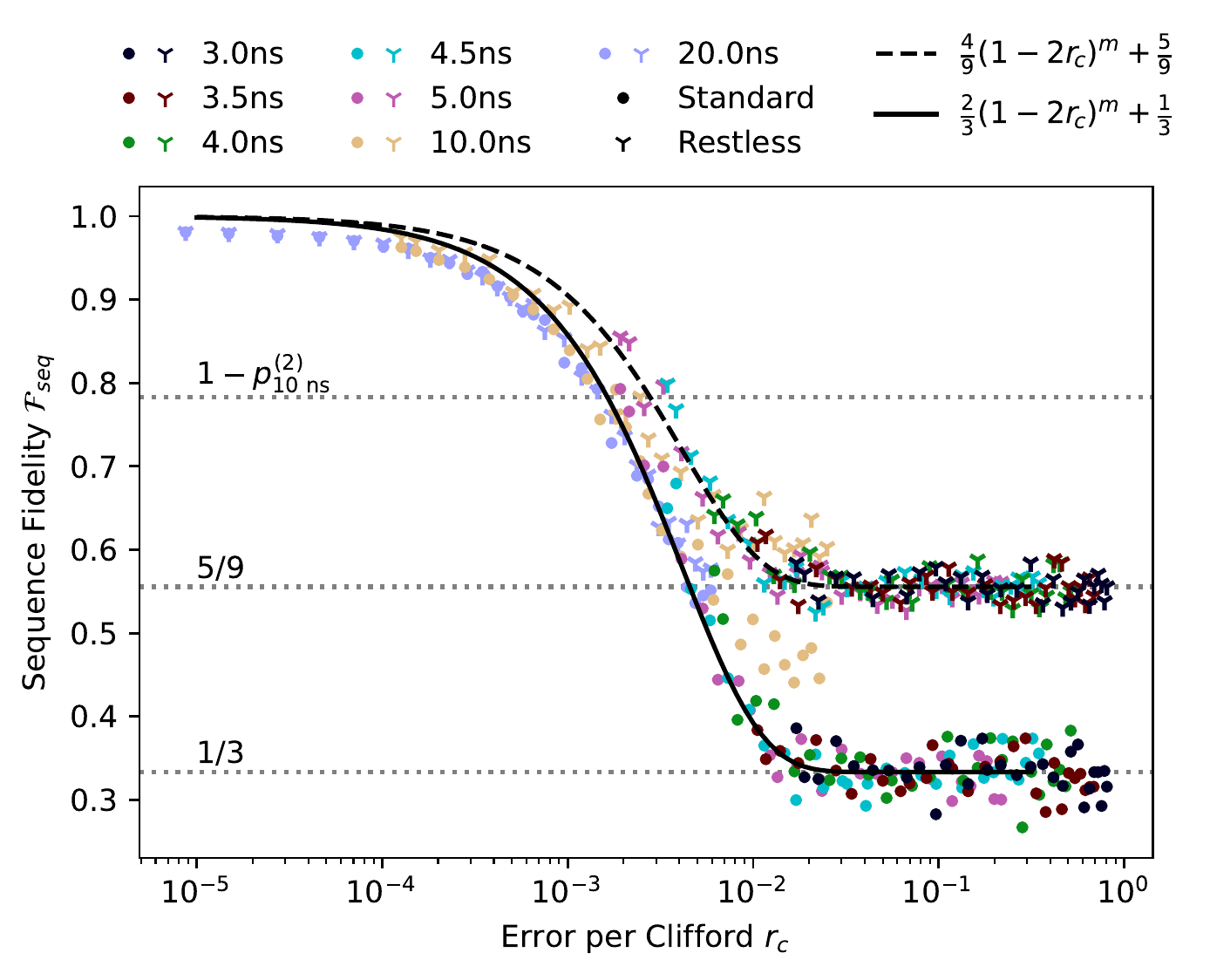}
    \caption{
    ORBIT cost function $\mathcal{F}_\text{seq}$ at $m=120$ versus Error per Clifford $r_c$ for various pulse durations, with an amplitude damping channel.
    Each point is the result of a single ORBIT experiment for a given pulse duration, DRAG parameter $\beta$, and circuit execution method.
    The lines are the ideal qubit sequence fidelities obtained from fully depolarized qutrit states without SPAM error.
    The $10$ and $20~\mathrm{ns}$ pulses do not extend across the full $r_c$ range since their narrow bandwidth implies a lower leakage than the high bandwidth $4.5~\mathrm{ns}$ and shorter pulses.
    }
    \label{fig:orbit}
\end{figure}
As expected, we observe a decrease in the ORBIT sequence fidelity as $r_c$ increases~\cite{Mckay2019}, see Fig.~\ref{fig:orbit}.
Crucially, we do not observe a change in the functional form of the relationship between $\mathcal{F}_\text{seq}$ and $r_c$ with restless circuit execution, both cases follow Eq.~(\ref{eq:rb_ideal}). 
Interestingly, we find that the high-levels of accumulated leakage occurring with short pulse durations, i.e., $<5~\mathrm{ns}$, impact the level at which $\mathcal{F}_\text{seq}$ settles.
Standard and restless circuit execution settle at $B=1/3$ and $B=5/9$, respectively, see Fig.~\ref{fig:orbit}.
We would naively assume that the sequence fidelity settles at $1/3$ for high infidelity leaky Clifford gates, as the circuit is equivalent to a fully depolarizing channel in the qutrit basis.
However, this is only the case for standard circuit execution and not for restless circuit execution.
This difference is a result of the restless post-processing.

Consider $K$ single-qutrit ORBIT circuits, constructed from high-leakage gates, that are sufficiently deep to fully depolarize the qutrit, i.e., the state immediately prior to measurement is $\rho_\text{d}=\frac{1}{3}\sum_{i=0}^2\ket{i}\!\bra{i}$.
Therefore, the measured outcomes do not depend on the initial state.
Our measurement outcome probabilities are thus the same regardless of the circuit execution method.
With standard circuit execution, the sequence fidelity is the probability to measure $\ket{0}$, which for $\rho_\text{d}$ is $1/3$.
In the absence of SPAM errors, the fit parameters are $A=2/3$ and $B=1/3$ given estimates of the boundary conditions at $m=0$ and $m\rightarrow\infty$ from Eq.~(\ref{eq:rb_ideal}).
By contrast, restless post-processing computes $\mathcal{F}_\text{seq}$ as the probability that the measurement outcome $M_{kj}$ is the same as the previous outcome $M_{k-1,j}$~\cite{Tornow2022}, i.e.,
\begin{equation}
    \begin{aligned}\label{eqn:f_seq_restless}
        \mathcal{F}_\text{seq}^{(\text{restless})} & = {\rm Pr}[M_{kj}=M_{k-1,j}] \\
        & = {\rm Pr}[M_{kj}=\texttt{`0'}]{\rm Pr}[M_{k-1,j}=\texttt{`0'}] \\
        & + {\rm Pr}[M_{kj}\neq{}\texttt{`0'}]{\rm Pr}[M_{k-1,j}\neq{}\texttt{`0'}] \\
        \left.\mathcal{F}_\text{seq}^{(\text{restless})}\right\vert_{r_c\to1}& = {\left(\frac{1}{3}\right)}^2 + {\left(1 - \frac{1}{3}\right)}^2 = \frac{5}{9}.
    \end{aligned}
\end{equation}
Note that the shots are time ordered: we interpret shot $M_{K,j}$ as $M_{0, j+1}$ owing to the sequence in which the circuits are run.
Equation~(\ref{eqn:f_seq_restless}) explains why restless measurements settle at $5/9$ in Fig.~\ref{fig:orbit} instead of $1/3$.
The inferred ideal fit parameters are thus $A_\text{leak.}=4/9$ and $B_\text{leak.}=5/9$.
These ``ideal'' sequence fidelities with and without restless post-processing are shown as black solid and dashed lines in Fig.~\ref{fig:orbit}.
An alternative to Clifford sequences that compose to the identity are sequences that compose to $X$.
These sequences result in different values of $A$ and $B$ and can be more sensitive to leakage, see Appendix~\ref{sec:compose_to_x}.

The insights gained on the ideal fit parameters $A$ and $B$ allow us to understand the sensitivity of ORBIT with restless circuit execution.
Kelly \emph{et al.} define the sensitivity of the ORBIT cost function as ${\rm d}\mathcal{F}_\text{seq}/{\rm d}r_c$~\cite{Kelly2014}.
For the single-qubit case, i.e., $\alpha=1-2r_c$, the sensitivity is $-2Am(1-2r_c)^{m-1}$.
The maximum sensitivity occurs at a sequence length $m^*$, defined in Ref.~\cite{Kelly2014}, and is given by ${\rm d}\mathcal{F}_\text{seq}/{\rm d}r_c\vert_{m^*}\approx -A/(er_c)$.
Therefore, with high leakage levels the restless ORBIT cost function is 33.3\% less sensitive than in the absence of leakage, i.e., $1-A_\text{leak.}/A$.

We hypothesise that sufficiently low accumulated leakage, during an ORBIT experiment, would recover the $1/2$ settling point for $\mathcal{F}_\text{seq}$ and sufficiently deep Clifford sequences ($m\gg{}1$) as the ORBIT gate sequences would result in a depolarizing channel for the qubit subspace instead of the qutrit space.
If the levels of accumulated leakage are sufficiently low, the probability to measure \texttt{`2'} would be negligible and thus the probability to measure \texttt{`0'} would be $50\%$ by normalization.

\section{Discussion and Conclusion\label{sec:conclusion}}

Advanced control methods help us design better quantum gates~\cite{Baum2021}.
Furthermore, methods like restless circuit execution reduce the footprint of characterization and calibration tasks~\cite{Tornow2022}.
It is also crucial to understand how reliable these new methods are. 
In this work we present a restless circuit execution simulator as a Markov Chain to investigate how much leakage restless measurements tolerate.
Leakage affects both standard and restless calibration experiments with restless circuit execution being only slightly more sensitive to leakage.
For example, fine amplitude calibration experiments can tolerate 0.5\% and 1\% leakage in standard and restless execution, respectively.
Crucially, these large leakage levels may be present only at the beginning of a set of calibration experiments.

The high leakage levels we investigate are only present in short pulses.
Such pulses are often encountered when pushing for faster gates, as exemplified by Refs.~\cite{Zhu2021, Werninghaus2021}.
In Ref.~\cite{Werninghaus2021}, short single-qubit pulses with closed-loop restless optimal control are designed by starting from a DRAG pulse.
Here, it is important that the cost function is sensitive to changes in the pulse shape.
The analysis in Sec.~\ref{sec:rb} complements the experimental observations of Refs.~\cite{Rol2017, Werninghaus2021} that restless ORBIT can optimize pulses even under high leakage-levels.
Crucially, the observations of Sec.~\ref{sec:rb} show a 1/3 decrease in the sensitivity of ORBIT due to the restless circuit execution and leakage.

Ultimately, under typical operating conditions of superconducting qubit hardware leakage is minimal and does not impact restless characterization and calibration.
It is thus neither necessary to employ leakage reset nor employ discriminators that classify more than the first two levels of the transmon.
This is in stark contrast with error correction where leakage is detrimental to code performance.

\section{Acknowledgements}

The code for the restless simulator is available at  \url{https://github.com/eggerdj/restless-simulator}.
IBM, the IBM logo, and ibm.com are trademarks of International Business Machines Corp., registered in many jurisdictions worldwide. Other product and service names might be trademarks of IBM or other companies. The current list of IBM trademarks is available at \url{https://www.ibm.com/legal/copytrade}.

\appendix

\section{Qubit Discrimination\label{app:discrimination}}

Superconducting qubits are measured by probing a readout resonator dispersively coupled to the qubit~\cite{Krantz2019}.
The qubit state imparts a frequency shift on the resonator which changes the transmission and reflection properties of a prob signal.
This signal, when down-converted and digitized, results in a point in the complex plan (IQ plane).
Discriminators that map complex IQ points to labels of the transmon state are trained by individually preparing the states of the transmon.
Discriminators to classify the first four levels of the transmon have been created~\cite{Fischer2022, miao2022, Chen2023}.
However, by default, most quantum computers employ a 0/1 discriminator, as exemplified by IBM Quantum backends.

\begin{figure}[h!]
    \centering
    \includegraphics[width=0.85\columnwidth]{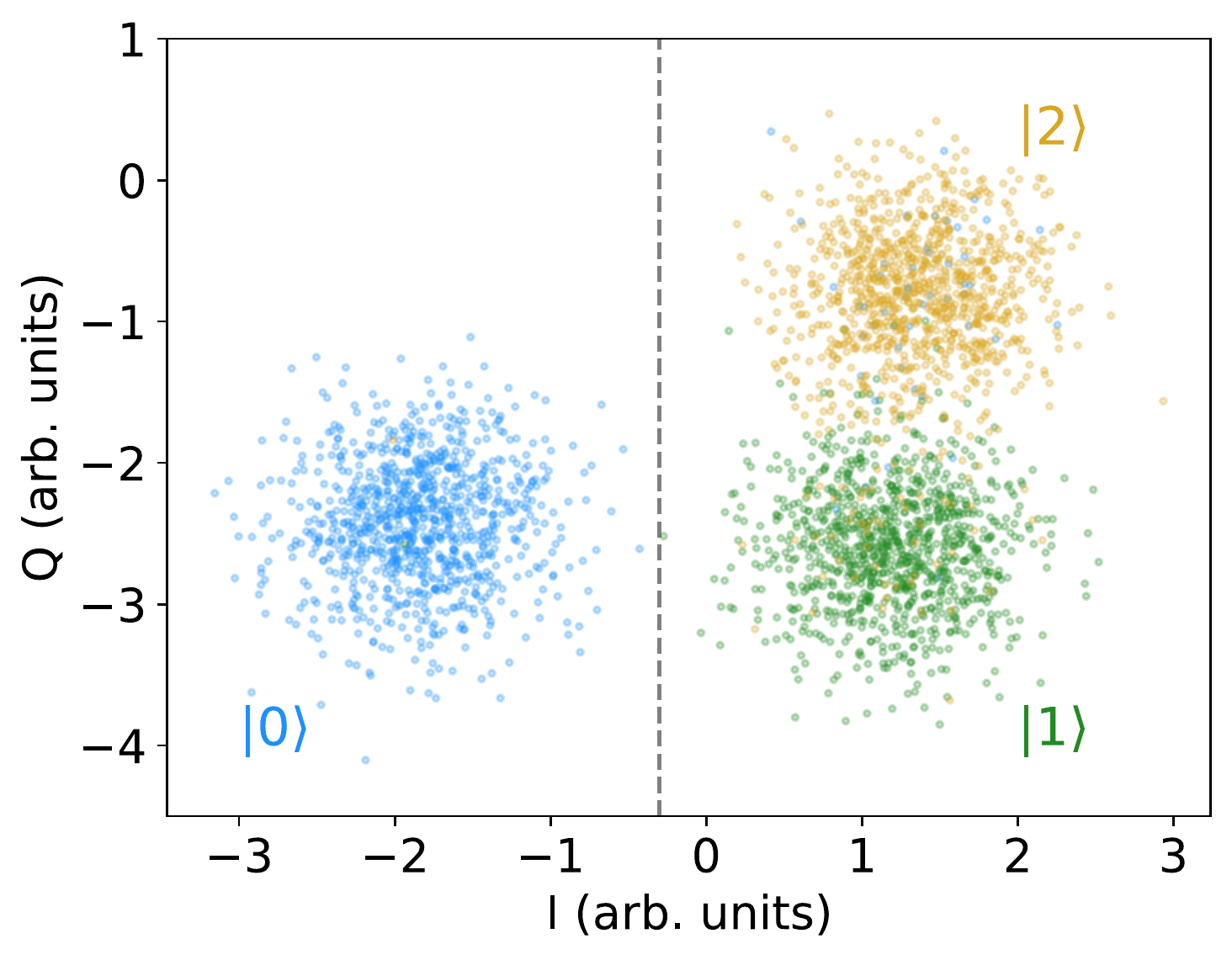}
    \caption{Example of a qubit discriminator classifying the first three levels of a transmon.
    The color of each shot indicates the state that is ideally prepared.
    The black dashed line shows a possible separation between the ground and higher excited states. Note that is is not the actual discriminator employed by the backend.
    }
    \label{fig:discriminator}
\end{figure}

Here, we illustrate how a qubit discrimination creates the measurement matrix given in Eq.~(\ref{eqn:meas_matrix}).
To this end we calibrate an $X$ gate between the $\ket{1}$ and $\ket{2}$ states of the transmon with a Rabi experiment implemented in Qiskit Experiments~\cite{QiskitExperiments}.
Next, we create three circuits that prepare the $\ket{0}$, $\ket{1}$, and $\ket{2}$ states.
We run these circuits with 1024 shots twice on an IBM Quantum backend: once requesting IQ points and once requesting classified data.
The IQ points show three clusters indicating that we successfully prepared the first three sates of the transmon, see Fig.~\ref{fig:discriminator}.
However, the classified data returns only \texttt{`0'} and \text{`1'} counts.
When the second excited state is prepared the backend identifies 1023 of the IQ points as \texttt{`1'}.
Therefore, up to SPAM errors, this discriminator classifies the transmon levels as $\ket{0}\to$\texttt{`0'}, $\ket{1}\to$\texttt{`1'}, $\ket{2}\to$\texttt{`1'}.
We model this classification with the matrix in Eq.~(\ref{eqn:meas_matrix}).

\section{Energy relaxation\label{app:damping}}

$T_1$ energy relaxation in superconducting transmon qutrits can be modelled as a sequential process between neighbouring states \cite{Fischer2022}.
States decay from $\ket{n}$ to $\ket{n-1}$ and the relaxation of $\ket{2}$ into $\ket{0}$ is suppressed.
A single-qutrit amplitude damping channel can be described in the operator-sum representation with four Kraus operators $K_i$, $i=\{0,1,2,3\}$~\cite{chessa2021} as
\begin{align}
\Lambda(\rho) = \sum_{i=0}^3{K_i\rho{}K_i^\dagger}
\end{align}
where
\begin{equation}
K_0 =
    \begin{bmatrix}
    1   & 0                 & 0                                     \\
    0   & \sqrt{1 - \Gamma_{0,1}} & 0                                     \\
    0   & 0                 & \sqrt{1 - \Gamma_{1,2} - \Gamma_{2,0}}   \\
    \end{bmatrix},
\end{equation}
\begin{equation}
K_1 =
    \begin{bmatrix}
    0   & \sqrt{\Gamma_{0,1}} & 0     \\
    0   & 0             & 0     \\
    0   & 0             & 0     \\
    \end{bmatrix},~\phantom{\text{and}}
\end{equation}
\begin{equation}
K_2 =
    \begin{bmatrix}
    0   & 0     & 0             \\
    0   & 0     & \sqrt{\Gamma_{1,2}} \\
    0   & 0     & 0             \\
    \end{bmatrix},~\text{and}
\end{equation}
\begin{equation}
K_3 =
    \begin{bmatrix}
    0   & 0     & \sqrt{\Gamma_{0,2}} \\
    0   & 0     & 0             \\
    0   & 0     & 0             \\
    \end{bmatrix}.~\phantom{\text{and}}
\end{equation}
The decay rates $\Gamma_{i,j}$ describe the strength of the decoherence path from $\ket{j}$ to $\ket{i}$. The sequential relaxation process for superconducting transmons implies that $\Gamma_{0,2}=0$.

To include amplitude damping in a qutrit circuit, we insert an error-channel instruction after each qutrit gate. 
The restless simulator, which leverages Qiskit~\cite{Qiskit2023}, supports unitary as well as completely positive trace-preserving maps. 
This allows us to model decoherence as a quantum circuit instruction that occurs after every gate.
The decay rates for the amplitude damping instructions are computed based on the pulse duration $\tau$ of the prior qutrit unitary gate and relaxation times for the first- and second-excited states of the qutrit.
We used relaxation times of $T_{0,1}=100~\rm{\mu{}s}$ and $T_{1,2}=73~\rm{\mu{}s}$ for the first- and second-excited states, respectively~\cite{Fischer2022}. 
Given the qutrit relaxation times, the decay rates are computed as
$\Gamma_{i,j} = 1 - \exp(-\tau/T_{i,j})$.

\section{ORBIT measurements as function of \emph{m}\label{app:orbit}}

\begin{figure*}[t]
    \centering
    \includegraphics[width=\linewidth,clip,trim=0 5 0 0]{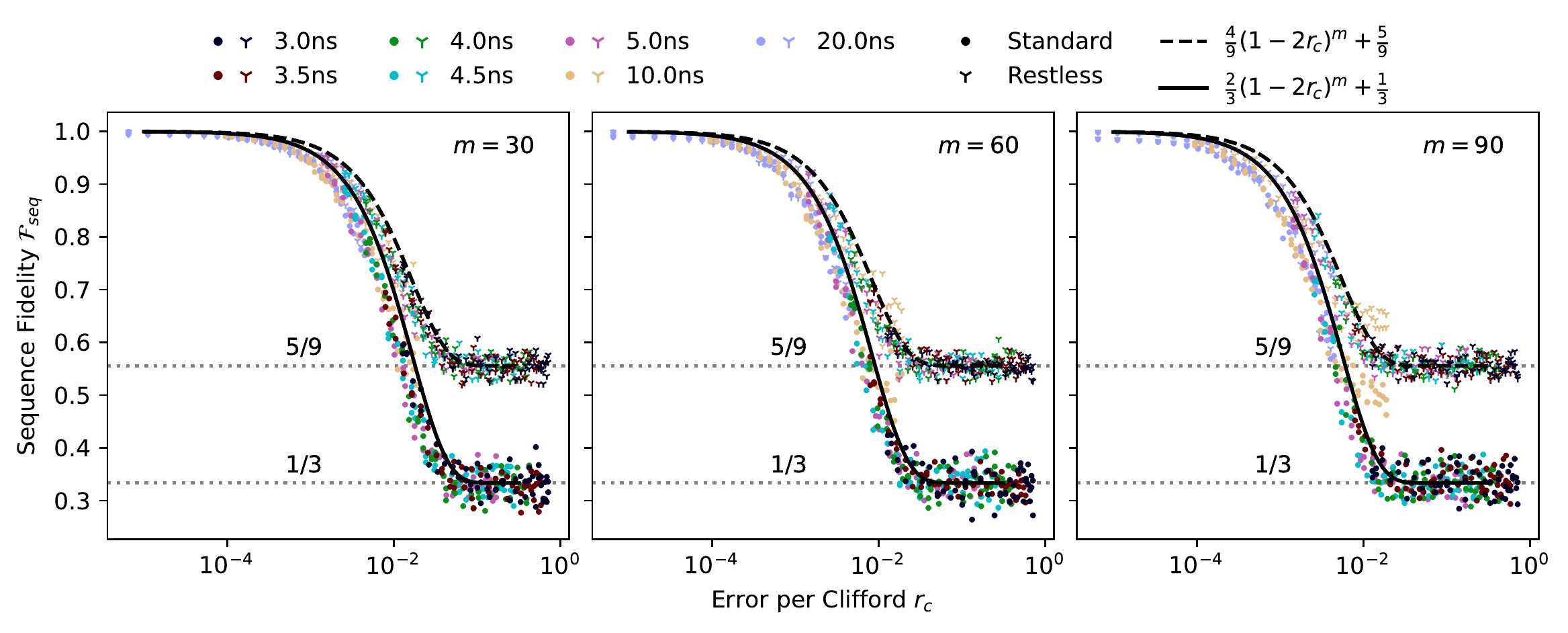}
    \caption{ORBIT cost function $\mathcal{F}_\text{seq}$ at $m=\{30, 60, 90\}$ versus Error per Clifford $r_c$ for various pulse durations, with an amplitude damping channel. 
    Each point is the result of a single ORBIT experiment for a given pulse duration, DRAG parameter $\beta$, and circuit execution method. 
    The lines are the ideal qubit sequence fidelities obtained from fully depolarized qutrit states without SPAM errors.}
    \label{fig:extra_orbit}
\end{figure*}

In addition to the results at $m=120$ of Fig.~\ref{fig:orbit} in the main text, we studied ORBIT sequence at depth $m\in\{30, 60, 90\}$.
These results were obtained in the same manner as those in the main text and confirm the same settling values at large $r_c$, see Fig.~\ref{fig:extra_orbit}.
As expected, as $m$ increases the value of $\mathcal{F}_\text{seq}(m)$ decreases thereby moving the curves in Fig.~\ref{fig:extra_orbit} towards the left.

\section{Clifford sequences that compose to \emph{X}\label{sec:compose_to_x}}

\begin{figure}
    \centering
    \includegraphics[width=\linewidth, clip,trim=7 0 10 0]{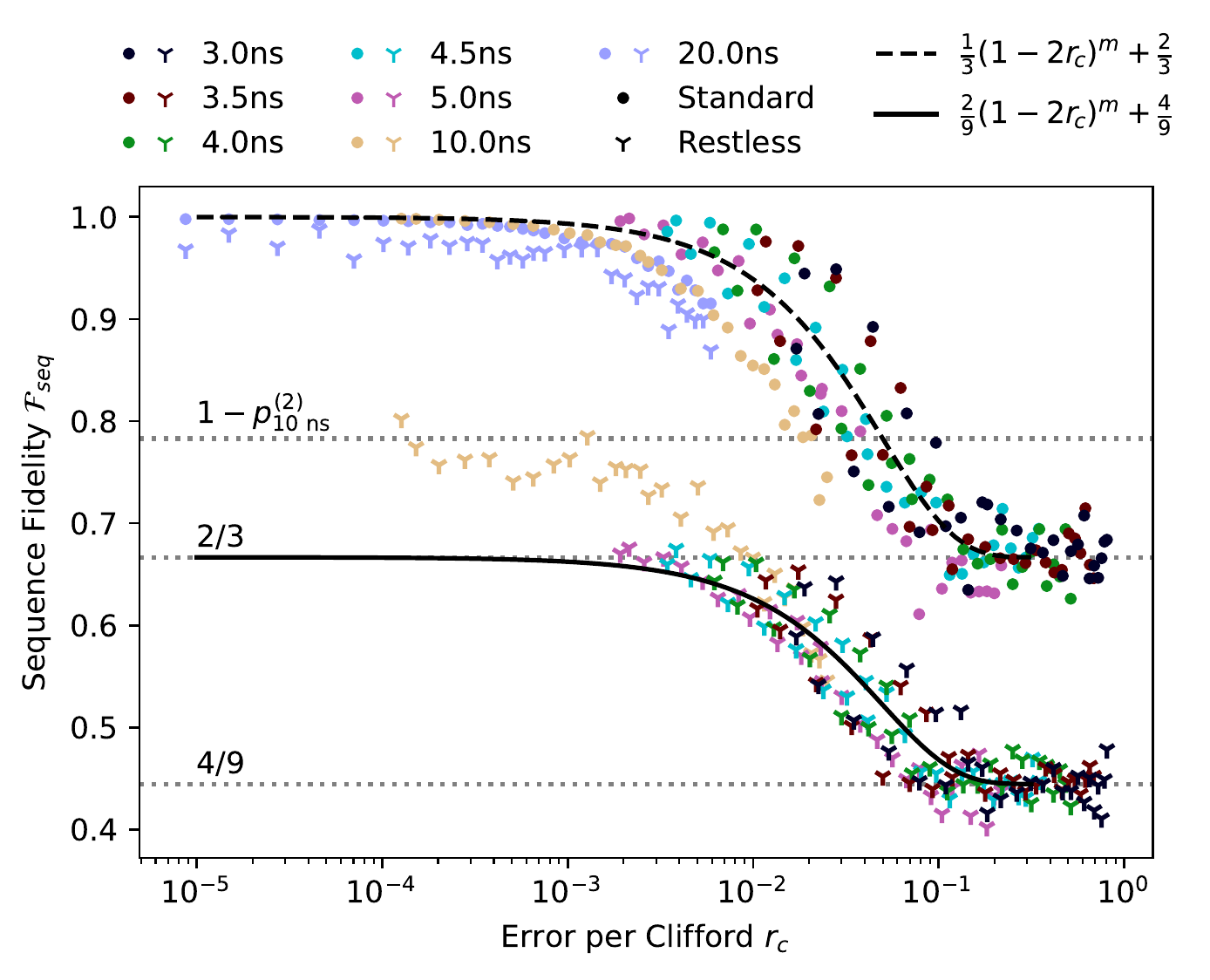}
    \caption{ORBIT cost function $\mathcal{F}_\text{seq}$ at $m=10$ versus Error per Clifford $r_c$ for various pulse durations, with an amplitude damping channel and a Clifford sequence that commutes to $X$ instead of the identity. 
    Each point is the result of a single ORBIT experiment with 1000 shots and $K=100$ instances of Clifford gate sequences for a given pulse duration, DRAG parameter $\beta$, and circuit execution method.
    The lines are the ideal qubit sequence fidelities obtained from fully depolarized qutrit states and the simulation results, without SPAM error.}
    \label{fig:orbit_10_commute_to_x}
\end{figure}

In Ref.~\cite{Rol2017}, Rol \emph{et al.} optimized DRAG pulses with Clifford sequences that compose to the $X$ gate.
We find that the settling values for ORBIT experiments, with standard and restless circuit execution, are different if the Clifford sequences compose to the $X$ gate instead of the identity.
Standard and restless measurements settle at $2/3$ and $4/9$, respectively, see Fig.~\ref{fig:orbit_10_commute_to_x}.
This differs from the $1/3$ and $5/9$ for compose-to-identity shown in Fig.~\ref{fig:orbit}.
Furthermore, restless ORBIT does not have a sequence fidelity close to one when the error per Clifford is low. 

An ORBIT experiment with Clifford sequences that compose to $X$ uses the probability to measure the excited state $\ket{1}$, instead of the ground-state $\ket{0}$, as the sequence fidelity.
Combining this with a fully-depolarized state and restless post-processing fully explains the different settling values seen in Fig.~\ref{fig:orbit_10_commute_to_x}.

For a fully depolarized state and a compose-to-$X$ Clifford sequence, the expected value at which standard circuit executed ORBIT experiments settle is ${\rm Pr}[M_{kj}=\texttt{`1'}]={\rm Pr}[\phi=\ket{1}] + {\rm Pr}[\phi=\ket{2}]=2/3$ since the discriminator classifies $\ket{2}$ as \texttt{`1'}.
The equivalent for restless circuit execution is
\begin{align}
    \mathcal{F}_\text{seq} & = {\rm Pr}[M_{kj}\neq{}M_{k-1,j}] \\ \notag
    & = {\rm Pr}[M_{kj}=\texttt{`0'}]{\rm Pr}[M_{k-1,j}=\texttt{`1'}] \\ \notag
    & + {\rm Pr}[M_{kj}=\texttt{`1'}]{\rm Pr}[M_{k-1,j}=\texttt{`0'}] \\ \notag
    & = 2\,{\rm Pr}[M_{kj}=\texttt{`0'}]{\rm Pr}[M_{k-1,j}=\texttt{`1'}] \\
    \left.\mathcal{F}_\text{seq}^{(\text{restless})}\right\vert_{r_c\to1}& = 2 \cdot \frac{1}{3}\cdot\frac{2}{3} = \frac{4}{9}.
\end{align}

The values of $\mathcal{F}_\text{seq}$ for low $r_c$ and compose-to-$X$ also differ from the results with compose-to-identity sequences in Fig.~\ref{fig:orbit}.
The accumulated leakage studied in Sec.~\ref{sec:leakage_buildup} explains this.
We assume that the accumulated levels of leakage in ORBIT are similar to those observed in the fine-amplitude calibration experiments.
For a given pulse duration $\tau$ restless circuit execution accumulates leakage over time and settles at $\smash{p_\tau^{(2)}}$.
We observe that the sequence fidelity for low $r_c$ pulses with restless circuit execution and a compose-to-$X$ Clifford sequence are less than one by approximately $\smash{p_\tau^{(2)}}$, i.e.,
\begin{equation}
    \left.\mathcal{F}^{(\tau)}_\text{seq}\right\vert_{r_c\to 0}\approx{}1 - p_\tau^{(2)}.
\end{equation}
For example, the population in the $\ket{2}$ state created by the $10~\rm{ns}$ pulse settles at $\smash{p_{10~\rm{ns}}^{(2)}}\approx{}0.217$ in Fig.~\ref{fig:fineamplitude_timedomain_leakage}. 
The maximum sequence fidelity $\smash{\mathcal{F}_\text{seq}=0.79\approx{}1 - p_{10~\rm{ns}}^{(2)}}$ coincides with the data in Fig.~\ref{fig:orbit_10_commute_to_x}.
These observations help us explain the offset from $\mathcal{F}_\text{seq}=1$ for $r_c\to 0$ in compose-to-$X$ Clifford sequences.

As the error per Clifford gate is low, we assume that the effect of the Clifford sequence is ``perfect'' in the qubit-subspace.
I.e., the probability to measure \texttt{`0'} and \texttt{`1'} conditioned on no leakage having occurred is the same as for an ideal $X$ gate without leakage.
Therefore, on average and once the leakage has settled as in Fig.~\ref{fig:fineamplitude_timedomain_leakage}, the two possible pre-measurement states are the fully-mixed states
\begin{equation}\notag
    \rho_X^{(a)} = \begin{bmatrix}
        1 - p^{(2)}_\tau & 0 & 0 \\
        0 & 0 & 0 \\
        0 & 0 & p^{(2)}_\tau \\
    \end{bmatrix}\,\,\text{and}\,\,
    \rho_X^{(b)} = \begin{bmatrix}
        0 & 0 & 0 \\
        0 & 1 - p^{(2)}_\tau & 0 \\
        0 & 0 & p^{(2)}_\tau \\
    \end{bmatrix}\!.
\end{equation}
Under compose-to-$X$ Clifford sequences, the qubit subspace alternates between $\ket{0}$ and $\ket{1}$ and thus the pre-measurement states alternate between $\smash{\rho_X^{(a)}}$ and $\smash{\rho_X^{(b)}}$.
Since the pulses with a low error per Clifford have a very low leakage, see Fig.~\ref{fig:optimized_pulse_performance}, if the qubit leaks into the $\ket{2}$ state it will stay there with high-probability.
For a compose-to-$X$ Clifford sequence, the pre-measurement states will alternate between $\rho_X^{(a)}$ and $\rho_X^{(b)}$.
Therefore, the restlessly post-processed sequence fidelity is the probability to have different measurement outcomes on both $\rho_X^{(a)}$ and $\rho_X^{(b)}$.
I.e., assuming that the state corresponding to circuit $k-1$ is $\rho^{(b)}_X$ then
\begin{align} \label{eq:restless_compose_x_prob} \notag
    \mathcal{F}_\text{seq} & = {\rm Pr}\left[M_{kj}=\texttt{`0'}|\rho_X^{(a)}\right]{\rm Pr}\left[M_{k-1,j}=\texttt{`1'}|\rho_X^{(b)}\right] \\ \notag
    & + {\rm Pr}\left[M_{kj}=\texttt{`1'}|\rho_X^{(a)}\right]{\rm Pr}\left[M_{k-1,j}=\texttt{`0'}|\rho_X^{(b)}\right]\\
    & = \left(1 - p^{(2)}_\tau\right) + 0 = 1 - p^{(2)}_\tau.
\end{align}
The same result is obtained if the state corresponding to a shot of circuit $k-1$ was $\smash{\rho_X^{(a)}}$ since the shot of circuit $k-1$ should then ideally correspond to $\smash{\rho_X^{(b)}}$.
If we use standard circuit execution, the qutrit is reset at each shot so that the state prior to measurement is $\smash{\rho_X^{(b)}}$. 
The sequence fidelity is thus one.
This explains why the restless sequence fidelities for low $r_c$ in Fig.~\ref{fig:orbit_10_commute_to_x} are lower than those we observed in Fig.~\ref{fig:fineamplitude_timedomain_leakage} by approximately $\smash{p^{(2)}_\tau}$ and why the sequence fidelities with standard circuit execution did not change.

As the fit-parameter $A$ for the compose-to-$X$ measurements is smaller than for the compose-to-identity measurements, the compose-to-$X$ sequences are less sensitive to variations in the error per Clifford.
If the accumulated leakage does not change during ORBIT optimization, then variations in the error per Clifford are dominated by the fidelity of the qubit-subspace operation.
If instead the infidelity of the gate is dominated by the level of accumulated leakage, then ORBIT with compose-to-$X$ sequences and restless is more sensitive than compose-to-identity sequences.
In fact, a \emph{short} compose-to-$X$ sequence, e.g., $m=10$, and restless circuit execution, may make a good ORBIT cost function to minimize small leakage levels and further optimize a high-fidelity gate.
By contrast, standard circuit execution may be less sensitive to small leakage amounts as the $\ket{2}$ state population is only a function of a single shot and not accumulated over time.

\section{Example Transition Matrices\label{app_tmats}}
Here, we give a few concrete examples of unitary matrices and the resulting transition matrices with which we studied leakage in restless measurements.
The fine-amplitude experiment shown in Fig.~\ref{fig:fineamplitude_timedomain_leakage} of the main text executes 17~circuits, each with an increasing number of leaky $X$ gates.
Each circuit $C_k$, with ${0\leq{}k<17}$, contains one ideal $\sqrt{X}$ gate followed by $k$ leaky $X$ gates.
Without decoherence, the effect of $C_k$ is
\begin{equation}\label{eq:circ_unitary_appendix}
      C_k(\ket{\nu}\!\!\bra{\nu})= U_X^kU_{\!\!\sqrt{\!X}}\ket{\nu}\!\!\bra{\nu}U_{\!\!\sqrt{\!X}}^\dagger{(U_X^\dagger)}^k.
\end{equation}
The leaky $X$~gate unitaries for the $5~\mathrm{ns}$ and $10~\mathrm{ns}$ duration pulses, as discussed in Sec.~\ref{sec:setup}, are computed from the time-order exponential of the Hamiltonian with a DRAG pulse which results in
\begin{widetext}
          \begin{align}
          U_X^{(5\,\mathrm{ns})} & = \begin{bmatrix}
              2.62\cdot{10}^{-2}-3.43\cdot{10}^{-4}j & 1.00 & -2.08\cdot{10}^{-2}+8.66\cdot{10}^{-2}j\\
              1.00-4.14\cdot{10}^{-5}j & -3.08\cdot{10}^{-2}+6.04\cdot{10}^{-3}j & 5.72\cdot{10}^{-2}+6.74\cdot{10}^{-2}j\\
              5.14\cdot{10}^{-2}-7.39\cdot{10}^{-2}j & -2.85\cdot{10}^{-2}-8.26\cdot{10}^{-2}j & -9.89\cdot{10}^{-1}+7.36\cdot{10}^{-2}j\\
            \end{bmatrix},~\text{and} \\
            U_X^{(10\,\mathrm{ns})} & = \begin{bmatrix}
              -7.33\cdot{10}^{-6}+8.05\cdot{10}^{-5}j & 1.00 & -10.00\cdot{10}^{-4}+1.33\cdot{10}^{-2}j\\
              1.00-5.86\cdot{10}^{-8}j & 7.68\cdot{10}^{-5}-8.44\cdot{10}^{-5}j & -1.19\cdot{10}^{-2}-6.10\cdot{10}^{-3}j\\
              -5.88\cdot{10}^{-3}+1.20\cdot{10}^{-2}j & -8.79\cdot{10}^{-3}-1.01\cdot{10}^{-2}j & -8.01\cdot{10}^{-1}+0.60j\\
            \end{bmatrix}. \label{eqn:umat}
        \end{align}
\end{widetext}
Note that the fidelity function in Eq.~(\ref{eqn:f_qpt}) which produced the underlying pulses is insensitive to the global phase.
Here, we display $U_X$ with a phase such that the $\ket{0}\!\!\bra{1}$ entry is real.
We compute the transition matrix $T_k$ for each circuit $C_k$ with Eq.~(\ref{eq:trans_mat_entries}) in the main text and Eqs.~(\ref{eq:circ_unitary_appendix}) to (\ref{eqn:umat}).
The resulting transition matrices $T_k^{(\tau)}$ for $\tau=5~\mathrm{ns}$ and $k=1$ and $16$ are
\begin{align}        \label{eq:trans_mat_fineamp_5ns_k1}
        T_1^{(5\,\mathrm{ns})} & = \begin{bmatrix}
          0.50 & 0.50 & 7.93\cdot{10}^{-3}\\
          0.50 & 0.49 & 7.81\cdot{10}^{-3}\\
          1.52\cdot{10}^{-3} & 1.42\cdot{10}^{-2} & 0.98\\
        \end{bmatrix},\\
        \label{eq:trans_mat_fineamp_5ns_k16}
        T_{16}^{(5\,\mathrm{ns})} & = \begin{bmatrix}
          0.34 & 0.43 & 0.23\\
          0.37 & 0.30 & 0.33\\
          0.29 & 0.27 & 0.44\\
        \end{bmatrix}.
\end{align}
The transition matrices for $10~\mathrm{ns}$ and $k=1$ and $16$ are
\begin{align}
\label{eq:trans_mat_fineamp_10ns_k1}
        T_1^{(10~\mathrm{ns})} & = \begin{bmatrix}
          0.50 & 0.50 & 1.79\cdot{10}^{-4}\\
          0.50 & 0.50 & 1.79\cdot{10}^{-4}\\
          3.44\cdot{10}^{-4} & 1.40\cdot{10}^{-5} & 1.00\\
        \end{bmatrix},\\
        \label{eq:trans_mat_fineamp_10ns_k16}
        T_{16}^{(10~\mathrm{ns})} & = \begin{bmatrix}
          0.50 & 0.50 & 1.53\cdot{10}^{-3}\\
          0.50 & 0.50 & 6.24\cdot{10}^{-4}\\
          1.08\cdot{10}^{-3} & 1.08\cdot{10}^{-3} & 1.00\\
        \end{bmatrix}.
    \end{align}
The column $i$ of the last row of a transition matrix is the probability to transition from state $\ket{i}$ to state $\ket{2}$.
As can be seen from the values in the bottom rows, multiple applications of a leaky $X$ gate accumulates leakage to a higher level than simply indicated by the standalone unitary, e.g., see Fig.~\ref{fig:optimized_pulse_performance}.
Furthermore, the longer $10~\mathrm{ns}$ pulse accumulates leakage at a much lower rate than the shorter $5~\mathrm{ns}$ pulse, which is substantiated by the simulation results discussed in Sec.~\ref{sec:leakage_buildup}.

\bibliography{references}

\end{document}